\documentclass[final,5p,times,twocolumn]{elsarticle}
\usepackage[colorlinks,citecolor=blue,linkcolor=blue,anchorcolor=blue,filecolor=blue,urlcolor=blue]{hyperref}
\usepackage{amssymb,amsmath}
\usepackage{CJKutf8}
\biboptions{sort&compress}

\journal{Physics Letters B}

\begin{document}
\begin{CJK*}{UTF8}{gbsn} 
\begin{frontmatter}


\title{Impact of tensor force on quantum shell effects in quasifission reactions}

\author{Liang Li~(李良)}
\address{School of Nuclear Science and Technology, University of Chinese Academy of Sciences, Beijing 100049, China}
\author{Lu Guo\corref{cor1}~(郭璐)}
\cortext[cor1]{Corresponding author}
\ead{luguo@ucas.ac.cn}
\address{School of Nuclear Science and Technology, University of Chinese Academy of Sciences, Beijing 100049, China}
\address{Institute of Theoretical Physics, Chinese Academy of Sciences, Beijing 100190, China}
\author{K. Godbey}
\address{Facility for Rare Isotope Beams, Michigan State University, East Lansing, Michigan 48824, USA}
\author{A. S. Umar}
\address{Department of Physics and Astronomy, Vanderbilt University, Nashville, Tennessee 37235, USA}

\begin{abstract}
Quantum shell effects drive many aspects of many-body quantal systems and their interactions. Among these are the
quasifission reactions that impede the formation of a compound nucleus in superheavy element (SHE) searches. Fragment production in quasifission is influenced by shell effects as a nontrivial manifestation of
microscopic dynamics hindering the full equilibration of the composite system to form the compound nucleus.
In this Letter, we use the microscopic time-dependent Hartree-Fock (TDHF) theory to study $^{48}$Ca+$^{249}$Bk collisions to
investigate the influence of the tensor component of the effective nucleon-nucleon interaction.
The results show that the inclusion of the tensor force causes the spherical shell effect to become more prominent, particularly for the neutron number yield whose peak is exactly at magic number $N=126$.
This suggests that the tensor force plays a compelling role in the evolution of dynamical shell effects in nuclear reactions, influencing the competition between spherical and deformed shell gaps.
\end{abstract}

\end{frontmatter}



In the description of bound fermionic quantal systems, the presence of shells carries a particular importance as they determine the degree of stability of that system.
This is especially relevant in nuclear physics, with protons and neutrons filling their respective shells with gaps between shells leading to the so-called magic numbers.
These closed-shell nuclei are much more tightly bound compared to the ones that are non-magic and thus extremely stable~\cite{myers1966,sobiczewski1966}.
The largest neutron shell closures
are believed to be at $N=126$ and possibly at $N=172$ or $N=184$, and for protons near $Z=114$ and feasibly
at higher values of $Z=120$, 124 or 126~\cite{cwiok1996,kruppa2000},
with efforts to extend and confirm these with searches for the island of stability.
In addition to the magic numbers of the spherical shells we also
see deformation induced shell closures such as the ones at $Z=36$~\cite{kozulin2022} and $Z=40$, and $Z=52$--56 observed in fission studies
that are believed to be associated with octupole deformed shapes in the outgoing fission fragments~\cite{scamps2018,scamps2019}.

The synthesis of superheavy elements (SHE) has been a stimulating topic in the forefront of nuclear science~\cite{dullmann2015}.
To date, the seventh row of the periodic table has been completed with the discovery of element $Z=118$,
Oganesson~\cite{oganessian2006}.
Despite this progress, it is still extremely difficult to produce SHE experimentally via cold or hot fusion reactions due to competing
reactions that hinder the formation of the superheavy compound nucleus. Most significant of these are the fusion-fission and
quasifission~\cite{vardaci2019} reactions.
Quasifission, as opposed to fusion, is a nonequilibrium process characterized by shorter interaction time, significant mass transfer, mass-angle correlations, and partial memory of the entrance channel~\cite{back1983,shen1987}.
Theoretical studies of quasifission are an indispensable aid to elucidate the underlying reaction mechanisms
due to the nontrivial interplay between the entrance and exit channels. Collision energy~\cite{nishio2012},
deformation and orientation of the collision partners~\cite{hinde1995},
neutron richness of the compound nucleus~\cite{hammerton2015}, and shell effects in the exit channel~\cite{wakhle2014} all
individually or collectively influence the quasifission dynamics. For example, an interesting phenomenon
of quasifission reactions with actinide targets is that a mass-asymmetric yield peak in the $^{208}$Pb mass region was observed at beam energies around the Coulomb barrier~\cite{gippner1986,itkis2004,kozulin2010}. In particular, the recent experiment simultaneously measured the atomic number and the mass of quasifission fragments and demonstrated the important role of the proton shell closure at $Z=82$ in the quasifission fragment production~\cite{morjean2017}.

A number of theoretical approaches have been developed that describe the quasifission in terms of multi-nucleon transfer
processes~\cite{adamian2003,zagrebaev2007,feng2009a,Dai2014,zhao2016,feng2017,zhang2018a,wu2019,wu2020,Jiang2020,zhu2021,wu2022,Sun2022_PRC105-054610}.
Recently, time-dependent Hartree-Fock (TDHF) theory has proven to be an excellent tool for studying quasifission dynamics~\cite{simenel2018,stevenson2019,godbey2020}.
In particular, mass-angle distributions and final fragment total kinetic energies (TKE)~\cite{kedziora2010,wakhle2014,oberacker2014,goddard2015,hammerton2015,umar2015a,umar2015c,umar2016,prasad2016,wang2016,sekizawa2016,yu2017,guo2018d,Li2019,godbey2019,sekizawa2017,sekizawa2019b} are in good agreement with
experimental observations. Furthermore,
TDHF studies of quasifission dynamics have taught us that dynamics itself may be dominated by shell effects~\cite{simenel2018,sekizawa2019}.
Despite the apparent strong differences between fission and quasifission, it is interesting to note that
similar shell effects are found in both mechanisms~\cite{godbey2019,scamps2018,scamps2019,simenel2021}.
One benefit of utilizing microscopic calculations is that their only inputs are the parameters of the energy density functional.
The parameters are usually fitted to nuclear structure properties only, consequently reaction calculations do not require additional parameters determined from the reaction mechanisms.
In TDHF applications, the Skyrme interaction is employed as the effective nucleon-nucleon interaction~\cite{skyrme1958} to construct the energy density functional. The full Skyrme interaction
includes central, spin-orbit, and tensor terms,
where the two-body tensor force is written as
\begin{align}
    v_{\mathrm{T}}&=\dfrac{t_\mathrm{e}}{2}\bigg\{\big[3({\sigma}_\mathrm{1}\cdot\mathbf{k}')({\sigma}_\mathrm{2}\cdot\mathbf{k}')-({\sigma}_\mathrm{1}\cdot{\sigma}_\mathrm{2})\mathbf{k}'^{\mathrm{2}}\big]\delta(\mathbf{r}_\mathrm{1}-\mathbf{r}_\mathrm{2}) \notag \\
    &+\delta(\mathbf{r}_\mathrm{1}-\mathbf{r}_\mathrm{2})\big[3({\sigma}_\mathrm{1}\cdot\mathbf{k})({\sigma}_\mathrm{2}\cdot\mathbf{k})-({\sigma}_\mathrm{1}\cdot{\sigma}_\mathrm{2})\mathbf{k}^\mathrm{2}\big]\bigg\} \notag \\
    &+t_\mathrm{o}\bigg\{3({\sigma}_\mathrm{1}\cdot\mathbf{k}')\delta(\mathbf{r}_\mathrm{1}-\mathbf{r}_\mathrm{2})({\sigma}_\mathrm{2}\cdot\mathbf{k})\notag \\
    &-({\sigma}_\mathrm{1}\cdot{\sigma}_\mathrm{2})\mathbf{k}'
    \delta(\mathbf{r}_\mathrm{1}-\mathbf{r}_\mathrm{2})\mathbf{k}\bigg\}.
\end{align}
The coupling constants $t_\textrm{e}$ and $t_\textrm{o}$ represent the strengths of triplet-even and
triplet-odd tensor interactions, respectively.

Historically, the full Skyrme interaction was often simplified due to computational cost.
Early studies did not include the spin-orbit interaction, which led to the anomalies in fusion
studies that was resolved by the inclusion of the spin-orbit terms~\cite{Umar1986a,umar1989}.
Until recently, most calculations in nuclear dynamics omitted the tensor part of the effective interaction. Only in the last few years, several studies showed the effect of tensor force on fusion barriers and cross sections~\cite{Dai2014a,stevenson2016,guo2018,guo2018b,godbey2019c,sun2022}.
The tensor force used in most dynamical calculations has been the SLy5t interaction~\cite{colo2007}, which includes the tensor component to the previously introduced SLy5 force~\cite{chabanat1998a}.
In a recent investigation on Ni isotopes~\cite{brink2018},
by comparing the calculations with and without tensor force, it has been shown that
the tensor part significantly affects the spin-orbit splitting of the proton $1f$ orbit
that may explain the survival of magicity far from the stability valley.
The influence of the tensor force was also extensively
studied in the superheavy region in Ref.~\cite{suckling2010}.

In this Letter, the first evidence for the impact of the tensor force on shell effects is investigated for quasifission reactions using the TDHF theory.

The TDHF equations governing the time-evolution of the system are obtained variationally
from a determinental many-body wavefunction, $\Phi(t)$, comprised of single-particle wave functions $\phi_{\lambda}$
\begin{equation}
    h(\{\phi_{\mu}\})\ \phi_{\lambda} (r,t) = i \hbar \frac{\partial}{\partial t} \phi_{\lambda} (r,t)
    \ \ \ \ (\lambda = 1,...,A) \ ,
    \label{eq:TDHF}
\end{equation}
where $h$ is the resulting single-particle Hamiltonian deduced from the effective interaction.
During the past decade it has become numerically feasible to perform TDHF calculations on a
3D Cartesian grid without any symmetry restrictions
and with much more accurate numerical methods~\cite{umar2006c,sekizawa2013,maruhn2014}.
\begin{figure}[h!]
\includegraphics[width=0.9\columnwidth]{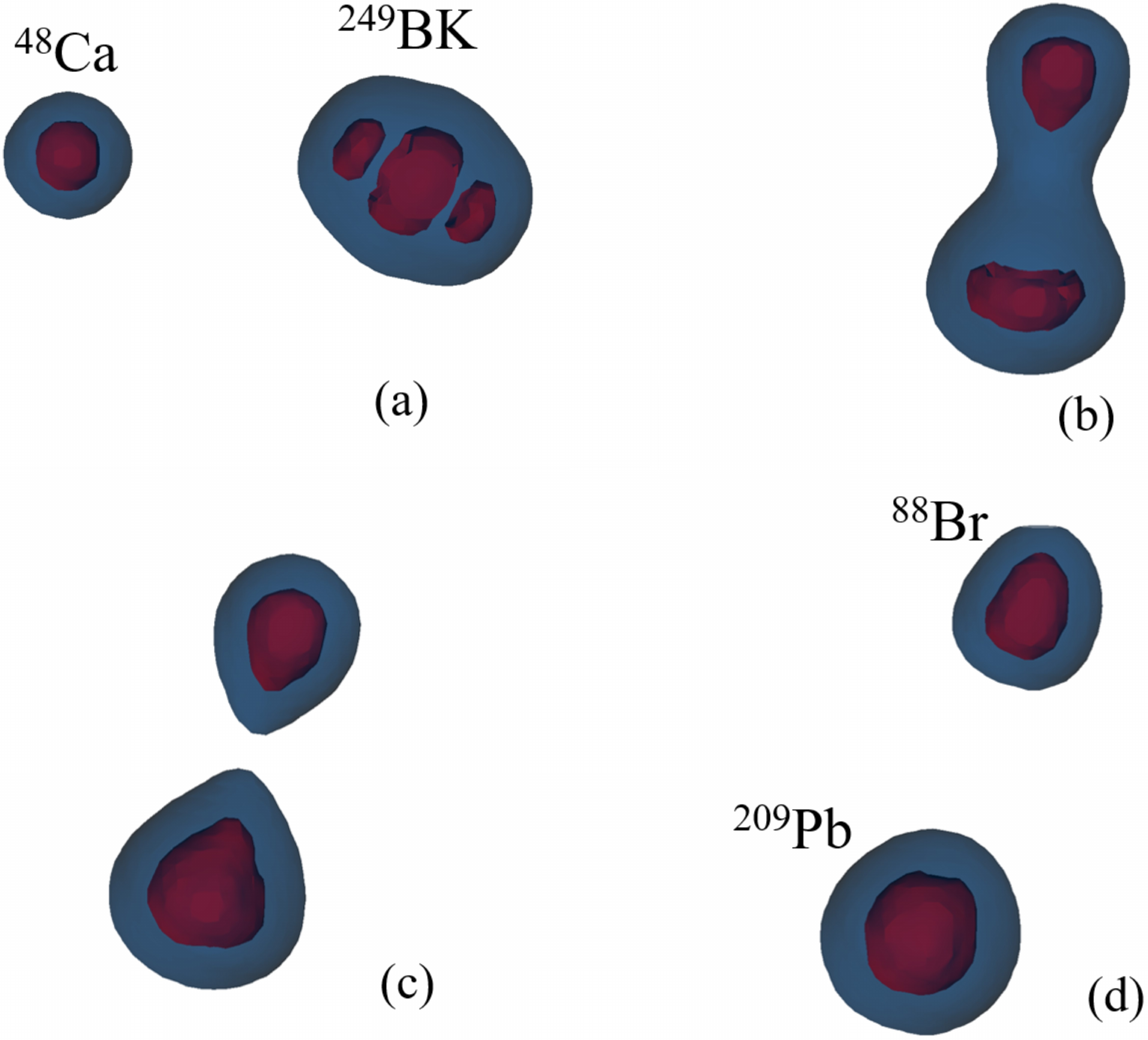}
\caption{Isodensity surfaces at $\rho = 0.03$, 0.151 $\mathrm{fm}^{-3}$ in midnight blue and red, respectively, for initial orientation $\beta= 30^\circ$ and impact parameter $b=2$~fm, shown at times (a) $t=0$, (b) 8.1, (c) 9.2, (d) 9.7~zs.}
\label{fig:density2}
\end{figure}
\begin{figure*}[t!]
\centering
\includegraphics*[width=1.5\columnwidth]{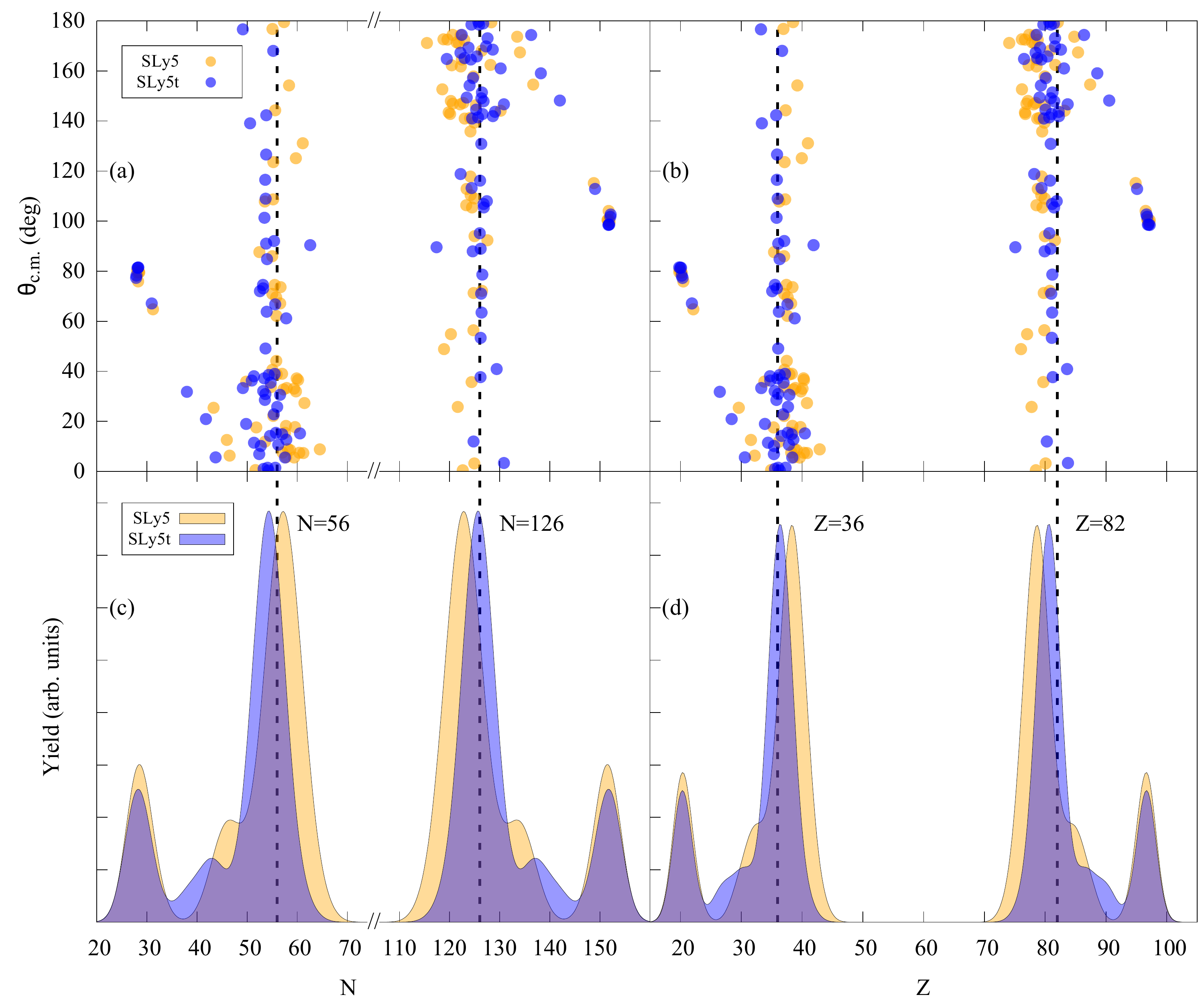}
\caption{Quasifission reaction for $^{48}$Ca+$^{249}$Bk at $E_{\mathrm{c.m.}}=234$~MeV. Distribution of scattering angle $\theta_{\mathrm{c.m.}}$ versus (a) neutron number $N$, (b) proton number $Z$ with (blue points) and without (orange points) tensor force. Fragment (c) neutron number yield, (d) proton number yield with (blue
shade) and without (orange shade) tensor force. The vertical dashed lines correspond to the spherical and deformed shell gaps at $N=56, 126$, and $Z=36, 82$, respectively.}
\label{fig:gaussn}
\end{figure*}

In this paper, we focus on quasifission in the reaction $^{48}\mathrm{Ca}+^{249}\mathrm{Bk}$.
The choice of this system was partially motivated by a recent extensive study of quasifission for the
same system using the SLy4d interaction~\cite{godbey2019}.
Static Hartree-Fock (HF) calculations result in a spherical
density distribution for $^{48}$Ca, while $^{249}$Bk shows prolate quadrupole
and hexadecupole deformation, in agreement with experimentally observed properties.
In practice, TDHF calculations are performed by placing the two static HF solutions obtained for the
target and projectile at a distance of 28~fm apart assuming that the two nuclei reached this
separation on a Coulomb trajectory and the relevant velocities are used to boost the nuclei. The
TDHF equations then propagate the nuclei in time as the reaction proceeds. In the case of quasifission
the time evolution stops when the final fragments reach a separation of about 28~fm. The final
scattering angle is then computed by extending the evolution to infinity on a Coulomb trajectory.
For the dynamical evolution, we use a numerical box of $60\times28\times48$~fm$^{3}$.
The time step is set to be $0.2~\mathrm{fm}/c$ and grid spacing is set to be 1.0~fm.
For the placement of the prolate deformed $^{249}$Bk nucleus, after the ground state wave function is
calculated we apply a rotation operator to the ground state using the b-spline interpolation method~\cite{pigg2014} to obtain wave functions of $^{249}$Bk for different orientations
with respect to the collision axis.
We have considered the orientations of $^{249}$Bk for $\beta=0^\circ$, ${30^\circ ,45^\circ ,60^\circ , 90^\circ }$ in the reaction plane. For each orientation calculations start at impact parameter b$=$ 0~fm with a step $\Delta$b=0.5~fm until the quasi-elastic reactions occur.
One such collision is depicted in Fig.~\ref{fig:density2} corresponding to initial conditions
$b=2$~fm and $\beta= 30^\circ$, shown at times (a) $t=0$, (b) 8.1, (c) 9.2, (d) 9.7~zs. The light and heavy fragments are $^{88}$Br and $^{209}$Pb, respectively.

Depending on the contact time each quasifission run takes about 1.5-4 days on a 16 processors modern workstation using
all processors in parallel.
\begin{figure}[t!]
\includegraphics*[width=8.6cm]{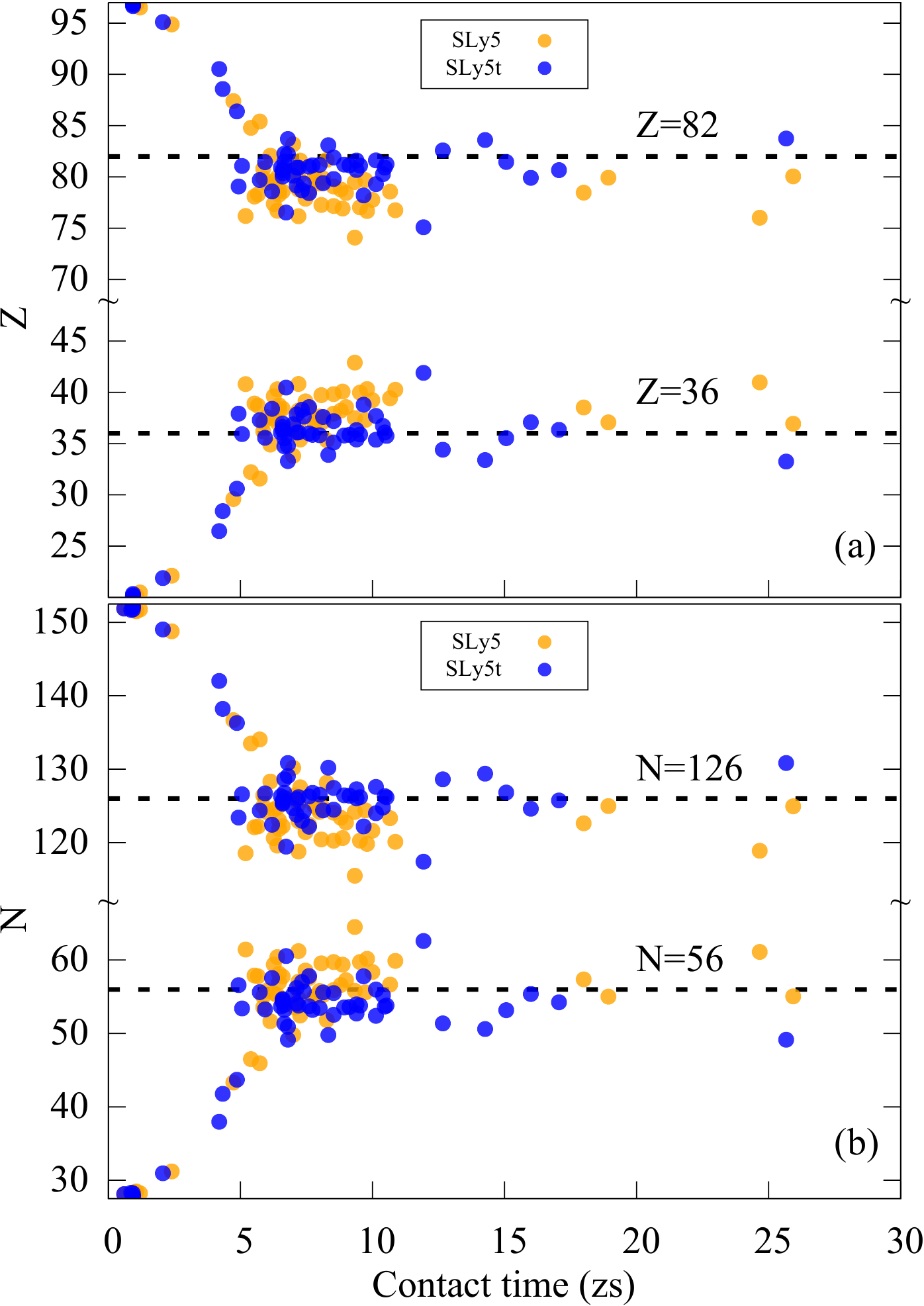}
\caption{(a) Proton number $Z$, (b) neutron number $N$ of quasifission fragments versus contact time with (blue points) and without (orange points) tensor force. The horizontal dashed lines represent spherical and deformed shell gaps.}
\label{fig:nztime}
\end{figure}
Results from different impact parameters and orientations have different contributions to the yield. We count the contributions using
\begin{equation}\label{3.1}
\sigma_{\lambda} \propto \int_{b_{\mathrm{min}}}^{b_{\mathrm{max}}} b d b \int_{0}^{\frac{\pi}{2}} d \beta \sin \beta P_{b}^{(\lambda)}(\beta),
\end{equation}
where  $\lambda$ represents a specific reaction channel, $P_{b}^{(\lambda)}(\beta)$ is the probability for a given impact parameter $b$ and orientation $\beta$. Its value is $0$ or $1$ for the reaction channel $\lambda$.

Each point in Fig.~\ref{fig:gaussn}(a) represents a TDHF result for reaction $^{48}$Ca+$^{249}$Bk at an impact parameter $b$ and orientation $\beta$.
The orange and blue points in Fig.~\ref{fig:gaussn}(a)--(c) represent TDHF results using SLy5
and SLy5t Skyrme parameter sets. The only difference of the two sets is that SLy5t includes the tensor force term of the Skyrme effective interaction and SLy5 does not~\cite{chabanat1998a,colo2007}. The results of the two sets are then compared to address the change that is caused by the tensor force. Figure~\ref{fig:gaussn}(a) shows the correlations between scattering angle $\theta_{\mathrm{c.m.}}$ and neutron number $N$. We see more blue points lie on the vertical $N=126$ dashed line than orange points. Also, most orange points are on the left side of the $N=126$ line, except for the several rightmost points. This suggests that quasifission reactions with tensor force favor production of more fragments with neutron magic number $N=126$. The neutron number of the rightmost points is $N=152$, the same as the target $^{249}$Bk, corresponding to the trajectories where quasi-elastic reactions occur.
With all the results from different orientations and impact parameters, it is possible to get smooth distributions of neutron and proton numbers in Fig.~\ref{fig:gaussn}(c,d).
In Fig.~\ref{fig:gaussn}(c), the corresponding neutron number yield is plotted with (blue shade) and without (orange shade) tensor force. The peak of the blue shade is centered at $N = 126$ while the peak of the orange shade is closer to $N = 122$. The yield distribution makes it more clear that the maximum production of heavy fragments with tensor force is affected by the $N = 126$ shell closure. Fig.~\ref{fig:gaussn}(b) shows the correlations between scattering angle $\theta_{\mathrm{c.m.}}$ and proton number $Z$. Blue points are systematically closer to the $Z=82$ line than orange points. This is also mirrored in the positions of yield peaks in  Fig.~\ref{fig:gaussn}(d). The peak of the blue shade is very close to $Z=82$ line while the peak of the orange shade is around $Z=79$. This illustrates that $Z=82$ shell effects have greater impact on quasifission fragments with tensor force than those without tensor force.
\begin{figure}[t!]
\includegraphics*[scale=0.34]{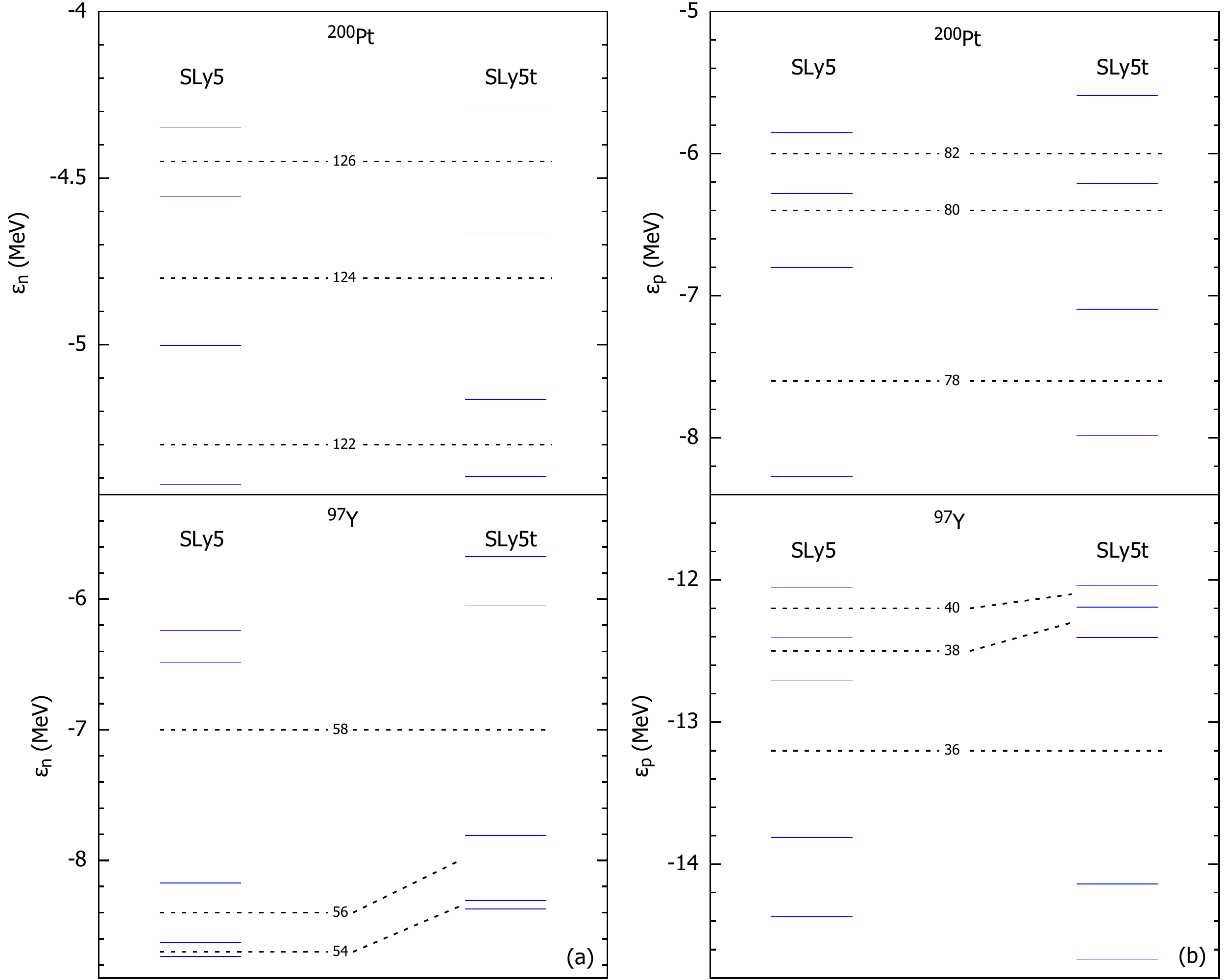}
\caption{Single-particle energy levels for the heavy ($^{200}$Pt) and light ($^{97}$Y)
 fragments with SLy5 and SLy5t forces for neutrons (a) and for protons (b).
 Please see the full discussion in text for details.}
\label{fig:spe}
\end{figure}

Figure~\ref{fig:nztime} shows the neutron and proton number of the heavy and light fragments as a
function of the contact time. Contact time is defined as the time elapsed from the first touching
of the target and projectile to the separation of the fragments (density overlap of about
$0.03$~fm$^{-3}$). We observe that the charge of the heavy fragments
with the tensor force (blue points) shifts towards the
$Z=82$ shell gap compared to the calculations without the tensor force (orange points).
This shift naturally impacts the charge of the light fragment, moving it from the deformed
shell gap of $Z=40$ (zirconium) to $Z=36$ (krypton).
As mentioned in the introduction this is
a manifestation of the octupole deformed shapes~\cite{scamps2018}
of the outgoing fragments.
Similarly, the neutron number of heavy fragments shift upwards to the $N=126$ shell closure
relative to the ones not using the tensor interaction. This pushes the neutron number of the
light fragment towards $N=54$ from the no-tensor value of $N=56$. Naturally, with a fixed
number of neutrons and protons in the system the competing shell gaps influence each other
for the formation of heavy and light fragments. From these we may conclude that spherical
shell gaps compete with the deformed ones for fragment formation and this competition is
sensitive to the details of the effective interaction employed. Finally, Fig.~\ref{fig:nztime}
also shows that the addition of the tensor force does not appreciably alter the contact time
of the reactions, most occurring between 5-10~zs, which may be considered fast quasifission.
Results using the SLy5 parameter set are consistent with Godbey~\textit{et al.}'s results using the SLy4d parameter set~\cite{godbey2019}. The SLy4d parameter set doesn't contain the tensor interaction, just as the SLy5 parameter set. Our results using the SLy5t parameter set show that, with the tensor force, the neutron shell effects play an important role and proton spherical shell effects become stronger. This has been observed experimentally with the spherical shell gap in quasifission for $^{48}$Ti+$^{238}$U~\cite{morjean2017} playing a vital role in fragment production.

The investigation of the shell effects at the single-particle level is complicated by the fact
that the quasifission studies presented in the manuscript involve a
distribution of points due to impact parameter and orientation
angle variations. In some sense this resembles mass/charge
distributions in an experimental study of QF or asymmetric fission, where
the shell effects are deduced from the peaks of the distributions.
In that sense this is different than a static structure study of
gaps where for a given EDF one set of results exists or even from
the traditional fission barrier calculations as a function of a
deformation parameter for which one could clearly see the evolution
of the gap from a single calculation.
Dynamical evolution also has the problem that the single-particle
energies contain the excitation energy of the system, which makes
it hard to ascertain what the actual energy of the single-particle
levels is.
Despite of these difficulties we thought that one can investigate
the deformation induced shell effects at the single particle
level by removing the excitation energy from these levels.
This was done by picking one of the points close to the SLy5 peak,
taking the fragments after the scission point ($^{97}$Y and $^{200}$Pt),
and extracting the internal energies by keeping the shape
of the fragments fixed using quadrupole and octupole deformation constraints of $\beta_{20}=0.552, 0.142$ and $\beta_{30}=0.049, 0.038$ for $^{97}$Y and $^{200}$Pt, respectively. We have repeated the calculation for
the same fragments using the SLy5t force. This is shown in Fig.~\ref{fig:spe}
for neutrons and protons. We see that the effect of
the tensor force is to enhance the $N=126$ gap for neutrons in the
heavy fragment, while the neutron gaps for the light fragment
remain about the same other than a shift. For protons we see a
small enhancement of the $Z=82$ gap for the heavy fragment but a much
larger enhancement of the $Z=36$ gap for the light fragment.
While a limited study, these observations are consistent with our
results.

The TDHF theory using the same Skyrme energy density functional with and without the tensor interaction was employed to simulate $^{48}$Ca+$^{249}$Bk collisions at a center-of-mass energy of 234~MeV. Calculations with five different orientations of the target have been combined to produce average fragment yield distributions. By comparing peaks of the yield distributions, we note that position of yield peaks moves towards magic number when the tensor force is included. Neutron-yield distributions in particular exhibit the most striking shift, with the yield peak moving to the neutron magic number $N = 126$. The differences caused by tensor force itself indicate that tensor force  strengthens spherical shell effects, resulting in more quasifission fragments produced whose neutron numbers are around $N=126$ and whose proton numbers are closer to $Z=82$. It is the first evidence that tensor force not only influences shell evolution in nuclear structure, but also plays a significant role in quasifission dynamics. This suggests further investigation on the role of tensor force in nuclear processes, as well as highlighting the delicate balance between spherical and deformed shell effects in nuclear dynamics.

This work has been supported by the Strategic Priority Research Program of Chinese Academy of
Sciences (Grant No. XDB34010000 and No. XDPB15), the National Natural Science Foundation
of China (Grants Nos. 11975237, 11575189, and 11790325), the U.S. Department of Energy under grant Nos.
DE-SC0013847 (Vanderbilt University) and DE-SC0013365 (Michigan State University). The computations in present work have been performed on the HPC cluster in Beijing PARATERA Tech Ltd.

\bibliographystyle{elsarticle-num}

\bibliography{VU_bibtex_master.bib}
\end{CJK*}
\end{document}